\documentclass[aps,%
reprint,
superscriptaddress,
amsmath,amssymb,floatfix,%pra,
prl
]{revtex4-2}
\usepackage{lipsum}
\usepackage{hhline}
\usepackage{ upgreek }

\usepackage{graphicx}% Include figure files
\usepackage{comment}
\usepackage{bm}% bold math
\usepackage{xcolor}
\usepackage{makecell}
\usepackage{amsmath,amsfonts,amssymb}
\DeclareMathAlphabet{\mathbbold}{U}{bbold}{m}{n}
\definecolor{darkblue}{rgb}{0,0,0.6}
\usepackage{hyperref}
\hypersetup{colorlinks,linkcolor=darkblue,citecolor=darkblue,urlcolor=darkblue}

\usepackage[normalem]{ulem}
\usepackage{tikz}
\usetikzlibrary{arrows.meta}

\begin{document}

%\title{The Geometry of an Intruder Governs its Dynamics in a Chiral Medium across Dilute and Hydrodynamic Regimes}

\title{Chiral Dynamics of an Intruder across Dilute and Hydrodynamic Regimes}

\author{Raphaël Maire}
\email{maire@ub.edu}
\affiliation{Departament de Física de la Mat\`eria Condensada, Universitat de Barcelona, C. Martí Franqu\`es 1, 08028 Barcelona, Spain}
\author{Ignacio Pagonabarraga}
\email{ipagonabarraga@ub.edu}
\affiliation{Departament de Física de la Mat\`eria Condensada, Universitat de Barcelona, C. Martí Franqu\`es 1, 08028 Barcelona, Spain}
\affiliation{University of Barcelona Institute of Complex Systems (UBICS), Universitat de Barcelona, 08028 Barcelona, Spain}

\date{\today}

\begin{abstract}
We introduce and simulate an analytically tractable model for an intruder of arbitrary shape in a nonequilibrium bath, with chirality originating from the bath, the intruder, or their coupling. In the dilute regime, a Langevin description derived from a Boltzmann-Lorentz equation shows how intruder geometry governs ratchet effects and odd response. In the dense regime, the dynamics of the intruder are instead governed by the hydrodynamic modes of the bath and edge currents, which are described by a Stokes equation including a chiral torque density. Our results link shape to chiral transport and show that odd response arises from distinct mechanisms in the dilute and dense limits.
\end{abstract}
\maketitle

\emph{Introduction}---The motion of an intruder immersed in a many-body medium is a classic problem in nonequilibrium physics. It provides a paradigmatic setting for coarse-graining, linking microscopic dynamics to effective equations of motion. For this reason, it has served as a testing ground for linear response~\cite{marconi2008fluctuation}, projection-based approaches~\cite{zwanzig2001nonequilibrium}, non-Gaussian fluctuations~\cite{van1986brownian}, non-Markovian effects~\cite{van1992stochastic}, long-time tails~\cite{widom1971velocity}, and broader fundamental~\cite{costantini2007granular,plyukhin2026langevin,plyukhin2006does} and applied questions~\cite{hanggi2009artificial}.

Compared with early studies of intruders in equilibrium media, where fluctuations are purely thermal, nonequilibrium environments such as granular gases display qualitatively new behavior, including spontaneous ratchet effects~\cite{costantini2007granular, eshuis2010experimental, gnoli2013brownian, plati2022collective, plati2019dynamical,manacorda2014coulomb,sarracino2013ratchet} and other pronounced signatures of broken detailed balance~\cite{sarracino2025nonequilibrium,talbot2011kinetics, sarracino2010granular, lucente2023revealing, Mendez_Garzo_2026, q1vm-13xl}. Whether such nonequilibrium features survive coarse-graining is not obvious \textit{a priori}~\cite{costantini2007granular, sarracino2010granular,puglisi2009irreversible}. A similar perspective has become central in active matter, where inclusions move through intrinsically driven biological or synthetic environments~\cite{marchetti2013hydrodynamics}, leading to transport phenomena similar to those found in granular gases~\cite{angelani2011active,reichhardt2017ratchet,krishnamurthy2016micrometre,anand2024transport,metzger2025exceptions,bechinger2016active,guidobaldi2014geometrical,ai2014transport, shea2024force, uchida2026designing, alston2026stochastic, benois2023enhanced, granek2022anomalous,demery2011perturbative,dean2011diffusion, granek2024colloquium}.

Chiral active matter is a particularly rich nonequilibrium setting~\cite{fruchart2023odd, liebchen2022chiral, lowen2016chirality,levis2019activity,digregorio2025phase,guo2026tuning,caprini2025Bubble,huang2021circular,musacchio2026circling,tan2022odd,siebers2023exploiting,caprini2025spontaneous,kuroda2023microscopic,kuroda2025singular, maitra2019spontaneous, kuroda2026designing, burekovic2026bulk}, with direct biological relevance~\cite{chen2025chirality, yashunsky2022chiral,inaki2016cell,juan2018myosin1d,di2011swimming, grober2023unconventional, nishiguchi2025vortex, villalobos2025active, takeuchi2026variousphasesactivematter}. More generally, mirror symmetry may be broken by the bath, by the intruder, or by their interaction, leading to odd transport or chirality-induced ratchets~\cite{Poggioli_Limmer_2023,kalz2022collisions,li2023chirality,Grober_Dhar_Saintillan_Palacci_2026}. Yet, a predictive microscopic theory connecting the geometry of an intruder to its effective chiral dynamics is still lacking. Recent works have obtained general Langevin descriptions for passive objects in chiral active baths and used symmetry to identify the allowed transport couplings~\cite{hargus2025odd,Passive2025HargusPRE}.  However, such an approach cannot determine which coupling coefficients are nonzero, how they explicitly depend on geometry, or whether a given effect originates, for example, from collective hydrodynamic effects or from Poissonian collisions.

To this end, our contributions in this Letter are threefold. (i) We disentangle the effects of shape chirality, bath-intruder chirality, and bath-bath chirality for the dynamics of the intruder. (ii) We derive the full chiral dynamics of an intruder of arbitrary shape in a dilute non-equilibrium bath and show that all forces, transport, and fluctuation coefficients are written as integrals characterizing the geometrical properties of the intruder. (iii) We distinguish this regime from the dense hydrodynamic one, where the bath behaves as a continuum whose response is dominated by correlated velocity profiles, and dilute collision physics plays only a secondary role. A longer companion paper contains all computations and additional details~\cite{long}.

\emph{Chiral Boltzmann-Lorentz---}We consider a closed two-dimensional intruder of mass $M$ and moment of inertia $I$. Its boundary is parametrized in the body frame by $s\in[0,1]$ as $\bm \rho_0(s)=(x_0(s),y_0(s))$. For example, a disk of radius $R$ corresponds to $\bm \rho_0(s)=(R\cos (2\pi s),R\sin (2\pi s))$. Choosing the body frame so that the intruder's center of mass is at the origin, the boundary in the laboratory frame is
\begin{equation}
    \bm r^{(c)}(s,t)=\bm X(t)+\bm r_0(s, t),\quad \bm r_0=\mathsf R\!\left(\varphi(t)\right)\cdot\bm \rho_0(s)
\end{equation}
where $\bm X(t)$ and $\varphi(t)$ denote the center-of-mass position and orientation of the intruder. The surrounding medium is a bath of point particles with mass $m$ at density $n_b$ interacting with the intruder only through instantaneous collisions at $\bm r^{(c)}$, which change the intruder velocities $\bm V$ and $\Omega$ and the particle velocity $\bm v$. Conservation of linear momentum imposes the following collision rule:
\begin{equation}
\bm V'=\bm V+\frac{\bm J}{M},\quad \Omega'=\Omega+\frac{\left(\bm r_0(s)\times \bm J\right)_z}{I},\quad \bm v'=\bm v-\frac{\bm J}{m},
\label{eq:collision_rule_main}
\end{equation}
where primes denote post-collisional quantities, $\bm J=J_n\hat{\bm n}+J_t\hat{\bm t}$ is the momentum change at collision, $\hat{\bm n}$ and $\hat{\bm t}=\hat{\bm z}\times\hat{\bm n}$ are the local normal and tangent at the contact point, and $\bm r_0$ is the lever arm. Defining the relative velocity at contact as $\bm g=\bm v-\bm V-\Omega \hat{\bm z}\times\bm r_0$, we denote by $g_n$ and $g_t$ its normal and tangential components. The details of the collision model are given in the companion paper; here, we present only the key formulas. Since the intruder is macroscopic, collisions with bath particles may be dissipative, which we encode in the normal sector as
\begin{equation}
g_n'=-\alpha g_n,
\label{eq:dissipation}
\end{equation}
where $\alpha$ is the restitution coefficient, with $\alpha<1$ for dissipative collisions. Chirality enters through the tangential sector via a transverse momentum transfer $J_t\neq 0$:
\begin{equation}
J_t=\frac{2\Delta}{\lambda_t},\quad \lambda_t=\frac1m+\frac1M+\frac{\kappa_t^2}{I},\quad \kappa_t=(\bm r_0\times\hat{\bm t})_z,
\label{eq:J_t_2}
\end{equation}
where $\lambda_t^{-1}$ is an effective mass, and the sign of the velocity scale $\Delta$ sets the handedness. This term can be interpreted as an effective contribution from bath-particle rotation, such as active spinners~\cite{eren2025collisional}, which appears as a tangential impulse upon collision with the intruder~\cite{long}. As in models with tangential forces~\cite{caporusso2024phase,caprini2025Bubble}, it injects energy and drives intruder rotation. Eqs.~\eqref{eq:dissipation} and \eqref{eq:J_t_2} fully specify the collision rule and the normal impulse:
\begin{equation}
    J_n=\frac{1+\alpha}{\lambda_n}g_n-\frac{\kappa_n\kappa_t}{I\lambda_n}J_t,\qquad \kappa_n=(\bm r_0\times\hat{\bm n})_z,
\end{equation}
with $\lambda_n=m^{-1}+M^{-1}+{\kappa_n^2}/{I}$.

In the dilute limit, where the bath mean free path exceeds the intruder size, the intruder weakly perturbs the bath. For a convex intruder, colliding particles are thus approximately sampled from the homogeneous velocity distribution $f(\bm v)$, which fully characterizes the bath. Since successive bath-intruder collisions are uncorrelated, the intruder probability density $P(\bm U,\bm Y,t)$, with $\bm Y=(\bm X,\varphi)$ and $\bm U=(\bm V,\Omega)$, obeys the Boltzmann-Lorentz equation
\begin{align}
\partial_t P+\bm U\cdot\bm \nabla_{\bm Y}P
=\int d&\bm U'[W(\bm U|\bm U',\bm Y)P(\bm U',\bm Y,t)\label{eq:BL_main}\nonumber\\
&-W(\bm U'|\bm U,\bm Y)P(\bm U,\bm Y,t)],
\end{align}
where $W(\bm U'|\bm U,\bm Y)$ is the collision-induced transition rate from $\bm U$ to $\bm U'$ at fixed $\bm Y$, whose explicit form follows from the collision rule and is given in the companion paper~\cite{long}. A van Kampen expansion of Eq.~\eqref{eq:BL_main} and a Gaussian closure then yields the Langevin equation~\cite{Meurs_Van2004,long}:
\begin{equation}
    \mathsf M\cdot \dot{\bm U}=\bm F-\Upgamma\cdot \bm U+\sqrt{\mathsf D}\cdot\bm \zeta,\qquad
    \dot{\bm Y}=\bm U,
    \label{eq:Langevin_main}
\end{equation}
with $\mathsf M=\rm{diag}(M,M,I)$, $\bm F=(F_x, F_y, F_\varphi)$ the generalized force array and $F_\varphi$ the torque, $\Upgamma$ the damping matrix, $\mathsf D$ a positive semi-definite matrix, and $\bm \zeta$ a Gaussian white noise of zero mean and unit variance. Eq.~\eqref{eq:Langevin_main} holds when the bath-particle inertia is small compared with both translational and rotational intruder inertia, $m/M\ll1$ and $mL^2/I\ll1$, where $L$ is the characteristic intruder size. Higher-order corrections generate nonlinearities and non-Gaussian noise~\cite{supp,PLYUKHIN2005198}, yielding nonequilibrium fluctuations beyond Eq.~\eqref{eq:Langevin_main}.

\begin{figure*}
    \centering
    \includegraphics[width = 0.9999\linewidth]{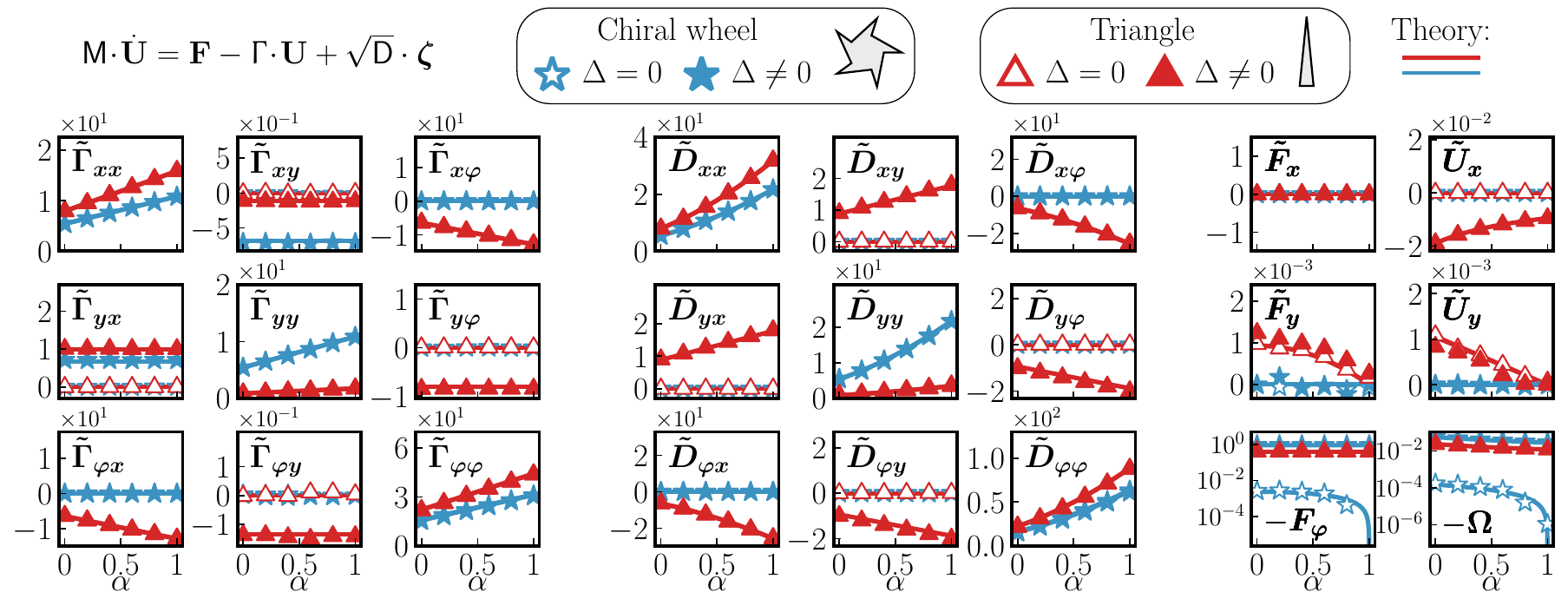}
    \caption{Parameters entering Eq.~\eqref{eq:Langevin_main} as functions of $\alpha$ for a chiral wheel and an isosceles triangle in a Gaussian bath. Simulation methods are detailed in the Supplemental Material (SM)~\cite{supp}, and the theory in the companion paper~\cite{long}. For the chiral wheel, the outermost-to-innermost radius ratio is $R/r=2$, and adjacent vertices are separated by angles $\pi(1\pm1)/5$. For the triangle, the height-to-base ratio is $h/a=5$. Units are set by length $\ell$ ($\ell=r$ or $a$), mass $m$, and time $\ell/\sqrt{T_b/m}$. We use $\Delta/\sqrt{T_b/m}=10^{-1}$ or $0$, $n_b\ell^2=1$, $M/m=300$, and the moment of inertia $I$ of a uniform mass distribution. Each point is an average over $100$ Gillespie simulations of $10^6$ collisions each. Tildes denote body-frame quantities, related to laboratory-frame quantities by rotation~\cite{long}.}
    \label{fig:coefficient}
\end{figure*}

Each coefficient in Eq.~\eqref{eq:Langevin_main} factorizes into a dimensional prefactor and a shape-dependent boundary integral. Table~\ref{tab:terms} lists these geometric integrals and the conditions under which the normal ($\alpha$) and chiral ($\Delta$) contributions are nonzero; complete prefactors are given in Ref.~\cite{long}. For example, at lowest order in the small-mass expansion, $\bm F=\bm F^{(\alpha)}+\bm F^{(\Delta)}$, where $\bm F^{(\alpha)}$ is independent of $\Delta$ and generated by the normal sector, while $\bm F^{(\Delta)}$ vanishes as $\Delta\to 0$. We will find that chiral intruder-bath collisions generate chiral dynamics for any shape, whereas normal non-equilibrium collisions do so only for a chiral intruder. Chirality in bath-bath collisions is irrelevant in the dilute limit because the bath enters only through $f(\bm v)$, which may be non-Gaussian but carries no handedness. Fig.~\ref{fig:coefficient} compares these predictions with direct simulations of Eq.~\eqref{eq:BL_main} for a chiral wheel and an elongated triangle, for $\Delta=0$ and $\Delta\neq0$. 

\begin{table}
\setcellgapes{2pt}
\makegapedcells
\begin{tabular}{|l||l|l|}
    \hline
%\multicolumn{3}{|c|}{\makecell{\textbf{Geometric structure of the effective couplings}\\
%$a,b\in\{x,y\}$ and $i,j\in\{x,y,\varphi\}$}} \\\hline
\makecell{}
&
\makecell{Normal component ($\alpha$)}
&
\makecell{Chiral component ($\vphantom{\dfrac 12}\Delta$) }
\\ \hhline{|=#=|=|}

\makecell{$\Gamma_{ab}$}
&
\makecell{$\propto Q_{ab}=\oint {ds}\hat n_a\hat n_b\neq 0$\\ Symmetric}
&
\makecell{$\propto (\bm \varepsilon \cdot \mathsf Q)_{ab}$}
\\ \hline

\makecell{$\Gamma_{\varphi a}$}
&
\makecell{$\propto \oint {ds} \kappa_n\hat n_a \neq 0\Rightarrow$\\ no $C_{n\geq 2}$ symmetry}
&
\makecell{$\propto \oint {ds}\kappa_t\hat n_a\neq 0\Rightarrow$\\ no $C_{n\geq 2}$ symmetry}
\\ \hline

\makecell{$\Gamma_{a \varphi}$}
&
\makecell{$\Gamma_{a\varphi}^{(\alpha)}=\Gamma_{\varphi a}^{(\alpha)}$}
&
\makecell{$\propto \oint {ds}\kappa_n\hat t_a\neq 0\Rightarrow$\\ no $C_{n\geq 2}$ symmetry}
\\ \hline

\makecell{$\Gamma_{\varphi \varphi}$}
&
\makecell{$\propto \oint {ds}\kappa_n^2>0$}
&
\makecell{$\propto \oint {ds}\kappa_n\kappa_t\neq 0$\\ $\Leftrightarrow$ Chiral shape}
\\ \hline

\makecell{$D_{ij}$}
&
\makecell{$2T_I \Gamma_{ij}^{(\alpha)}$}
&
\makecell{$(1+\alpha)T_b \big(\Gamma_{ij}^{(\Delta)}+\Gamma_{ji}^{(\Delta)}\big)$}
\\ \hline

\makecell{$F_{a}$}
&
\makecell{$\propto (T_b - T_I)\oint {ds}\kappa_n^2\hat{n}_a\neq 0$\\ $\Leftrightarrow$ Polar shape and\\non-equilibrium dynamics }
&
\makecell{0 when $m\Delta^2\ll T_b$}
\\ \hline

\makecell{$F_{\varphi}$}
&
\makecell{$\propto (T_b - T_I)\oint {ds}\kappa_n^3\neq 0$\\ $\Leftrightarrow$ Chiral shape and\\non-equilibrium dynamics }
&
\makecell{$\propto$ area $\mathcal A$ of the intru-\\der when $m\Delta^2\ll T_b$}
\\ \hline

\end{tabular}
\caption{Geometric structure of the couplings, with $\oint ds\equiv\int ds|\bm \rho_0'(s)|$ the line integral along the intruder boundary. The indices are: $a, b\in \{x, y\}$ and $i, j\in \{x, y, \varphi\}$.}
\label{tab:terms}
\vspace{-1.6em}
\end{table}
We begin with the damping matrix. As summarized in Table~\ref{tab:terms}, the translational block $\Gamma_{ab}^{(\alpha)}$ ($a, b\in \{x, y\}$) is symmetric, even for chiral shapes, whereas the chiral drive contribution $\Gamma_{ab}^{(\Delta)}$ contains an antisymmetric part, which is responsible for odd diffusivity~\cite{hargus2021odd} and in our case is proportional to the intruder perimeter $\mathcal P$, since $\bm \varepsilon\cdot \mathsf Q-(\bm \varepsilon\cdot \mathsf Q)^T=\mathcal P\bm\varepsilon$~\cite{long}, with $\bm \varepsilon$ the Levi-Civita pseudotensor. Consistently, in Fig.~\ref{fig:coefficient}, $\Gamma_{xy}$ is nonzero only for $\Delta\neq0$, even for the chiral wheel. Any $n$-fold rotational symmetry ($C_n$, $n\geq 2$) enforces $\Gamma_{a\varphi}=\Gamma_{\varphi a}=0$. However, the absence of such a symmetry is necessary, but not sufficient, for translation-rotation coupling. Accordingly, they vanish for the chiral wheel but not for the elongated triangle. In the normal sector, these couplings are reciprocal, $\Gamma_{a\varphi}^{(\alpha)}=\Gamma_{\varphi a}^{(\alpha)}$, even for chiral shapes, whereas explicit chiral interaction breaks this symmetry $\Gamma_{a\varphi}^{(\Delta)}\neq\Gamma_{\varphi a}^{(\Delta)}$. The chiral contribution to the rotational damping, $\Gamma_{\varphi\varphi}^{(\Delta)}$, is nonzero only for chiral shapes, since in our model the chiral impulse is the only source of transverse coupling and we exclude additional mechanisms such as rough contacts. We next consider the noise variance $\mathsf D$. %, which is symmetric because of our Markovian dynamics~\cite{PhysRevE.97.032117, han2021fluctuating}.
Its normal sector satisfies a fluctuation-dissipation relation $D_{ij}^{(\alpha)}=2T_I\Gamma_{ij}^{(\alpha)}$, with the intruder temperature $T_I$ instead of the bath temperature $T_b=m\int d\bm v \bm v^2/2 =m\langle \bm v^2\rangle/2$:
\begin{equation}
    T_I=\dfrac{1}{2} M\langle\bm V^2\rangle =I\langle \Omega^2\rangle =\frac{2}{3}\frac{\langle |\bm  v|^3\rangle}{\langle |\bm  v|^2\rangle\langle |\bm  v|\rangle}\dfrac{(1+\alpha)}{2}T_b,
    \label{eq:T_Imain}
\end{equation}
which reduces to $T_b$ at equilibrium, i.e., for elastic collisions ($\alpha=1$) and a Gaussian bath, for which the velocity-moment ratio equals unity. Eq.~\eqref{eq:T_Imain} holds for $m\Delta^2\ll T_b$, while for larger $\Delta$, translational and rotational degrees of freedom acquire distinct, generally orientation-dependent effective temperatures~\cite{long}. The chiral sector likewise obeys a symmetrized fluctuation-dissipation-like relation $D_{ij}^{(\Delta)}=(1+\alpha)T_b \big(\Gamma_{ij}^{(\Delta)}+\Gamma_{ji}^{(\Delta)}\big)$. 
Finally, the achiral sector generates fluctuation-driven ratchet terms $F_a^{(\alpha)}$ and $F_\varphi^{(\alpha)}$, proportional to the temperature difference between the intruder and the bath $T_b-T_I$~\cite{costantini2007granular}, for polar and chiral shapes, respectively. Thus, the triangle self-propels toward its apex, whereas the chiral wheel retains a finite torque even at $\Delta=0$. Both effects scale as $\mathcal O(m)$ in the small-bath-mass expansion and are therefore subleading to the $\Delta$-induced torque, which is not fluctuation-driven and scales as $\mathcal O(\sqrt{m})$. This hierarchy is confirmed in Fig.~\ref{fig:coefficient}. In short, we showed analytically how intruder geometry controls the allowed couplings in the dilute regime.

\begin{figure}
    \centering
    \includegraphics[width=0.99\linewidth]{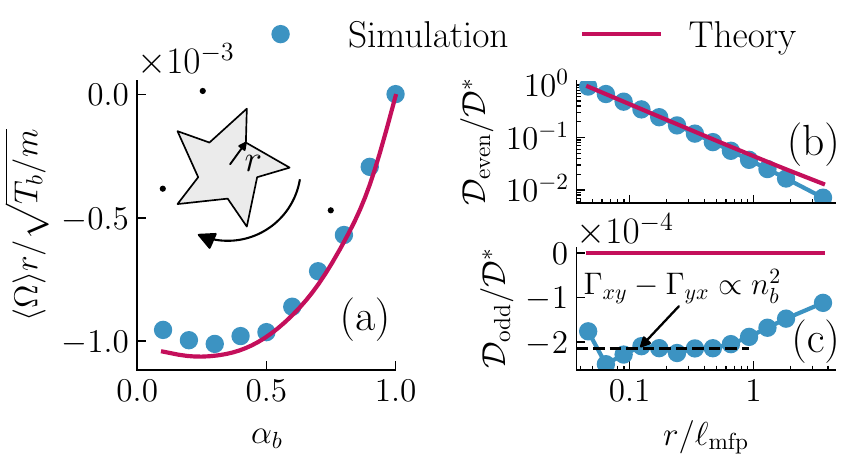}
    \caption{Intruder dynamics within an explicit bath. (a) Angular velocity versus bath restitution coefficient at $n_b r^2=10^{-3}$. The cartoon is to scale, with $r/\sigma=30$. (b), (c) Even and odd diffusivities: $\mathcal D=\mathcal D_{\rm even}\bm 1+\mathcal D_{\rm odd}\bm\varepsilon$, versus bath density for $\alpha_b=0.5$, with $\ell_{\rm mfp}$ the bath mean free path. Momentum is weakly relaxed by a Langevin bath, $\dot{\bm v}=-\gamma\bm v+\sqrt{2\gamma T_b/m}\bm\eta$, which prevents the divergence of $\mathcal D$ with system size~\cite{hansen2013theory}. Here, $\gamma^{-1}$ is about ten times the mean free-flight time. We define $\mathcal D^*=r\sqrt{T_b/m}$, tune $\Delta_b^\parallel$ to keep $T_b$ fixed for all $n_b$ and $\alpha_b$, and use $\alpha=1$, $\Delta=0$, $M/m=100$, and $I/Mr^2=1/9$. The geometry is set by $R/r=2$, with adjacent vertices separated by an angle $\pi(1\pm1/2)/5$. $\mathcal D$ is obtained via a Green-Kubo expression~\cite{Passive2025HargusPRE} and the measured $\Upgamma$ and $\mathsf D$. Data are averaged over at least 100 independent simulations of duration $3\times10^7r/\sqrt{T_b/m}$.}
    \label{fig:EDMD}
\end{figure}

\emph{Direct bath simulation---}To go beyond the dilute, homogeneous-bath limit, we simulate the full intruder-bath system by event-driven molecular dynamics~\cite{smallenburg2022efficient}, with bath particles of diameter $\sigma$; details are given in the SM~\cite{supp}. Our first result is that the dilute, homogeneous-bath prediction is recovered when the bath is modeled explicitly as a collection of interacting particles. However, because the bath is now modeled explicitly, we must specify its collision rule. We choose achiral bath--bath collisions with normal and tangential impulses
$J_n^{(b)}=\frac{m}{2}(1+\alpha_b)g_n-m\Delta_b^\parallel$ and $J_t^{(b)}=0$, where $\Delta_b^\parallel$ acts along the collision normal and sets $T_b(\alpha_b,\Delta_b^\parallel)$~\cite{brito2026dynamicpropertiescollisionalmodel,brito2013hydrodynamic}.
Even for elastic, achiral intruder-bath collisions ($\alpha=1$, $\Delta=0$), bath-bath dissipation ($\alpha_b<1$) makes the velocity distribution non-Gaussian, yielding $T_I\neq T_b$ and a finite torque on an intruder with a chiral shape. Fig.~\ref{fig:EDMD}(a) shows that, in the dilute regime, the measured angular velocity agrees with the theory obtained from the measured bath velocity moments through Eq.~\eqref{eq:T_Imain} and the analytical expression for $F_\varphi$~\cite{long}, even for this slightly nonconvex intruder. Thus, in the dilute, uncorrelated-collision limit, all bath information relevant to the intruder dynamics is encoded in its velocity distribution.

We next test the breakdown of the dilute theory at finite bath density. Figs.~\ref{fig:EDMD}(b) and (c) show the even and odd diffusivities, $\mathcal D_{\rm even}$ and $\mathcal D_{\rm odd}$, distinct from the noise matrix $\mathsf D$, as $n_b$ increases (equivalently the mean free path $\ell_{\rm mfp}(n_b)$ decreases). The theory predicts $\mathcal D_{\rm even}$ accurately down to $\ell_{\rm mfp}\sim r$. By contrast, $\mathcal D_{\rm odd}$ remains finite although Boltzmann-Lorentz theory predicts $\mathcal D_{\rm odd}=0$ for a chiral intruder in an achiral bath. This weak residual response approaches a plateau as $n_b\to0$ and can be shown to originate from the antisymmetric part of $\Upgamma$, which is found to scale as $n_b^2$, whereas its symmetric part scales as $n_b$. The quadratic scaling signals bath-intruder correlations beyond the Boltzmann-Lorentz approximation and identifies this odd response, unlike the ratchet torque in Fig.~\ref{fig:EDMD}(a), as genuinely collective. At $\ell_{\rm mfp}\lesssim r$, the prediction for $\mathcal D_{\rm even}$ also fails because the bath transports momentum and the drag crosses over from Epstein to Stokes behavior~\cite{stoyanovskaya2020simulations,masters1981molecular,cukier1980microscopic,scharf1970stokes}. This marks the crossover from Boltzmann-Lorentz kinetics to hydrodynamics.

\emph{Hydrodynamics---}We now turn to the hydrodynamic regime. A chiral intruder in an achiral nonequilibrium bath is subtle because the relevant effects are largely fluctuation-driven~\cite{jimenez2024fluctuations} with the observed $n_b^2$ scaling reflecting a feedback loop: fluctuating ratchet torques perturb the bath, which then acts back on the intruder. The converse case, an achiral intruder in a chiral fluid, is more tractable. To isolate bath chirality, which is irrelevant in the dilute limit, we set bath-intruder chirality to zero ($\Delta=0$) and introduce chirality only through bath-bath impulses, $J_n^{(b)}=\frac{m}{2}(1+\alpha_b)g_n$ and $J_t^{(b)}=m\Delta_b$. In the dense regime, $J_t^{(b)}$ generates a torque density $\tau(n_b,\alpha_b,\Delta_b)$~\cite{maire2026kinetic}, the antisymmetric counterpart of the pressure $p$ in the homogeneous stress $\mathsf \Sigma^{h}=-p\bm 1+\tau\bm\varepsilon$. Previous studies of 2D intruders focused mainly on odd viscosity, which can produce odd responses in compressible media~\cite{hosaka2021nonreciprocal,lier2023lift,daddi2026exact} or in incompressible media with slip boundaries~\cite{hosaka2023lorentz, lou2022odd, lier2024slip}. However, in the transverse-impulse model studied here, odd viscosity is found to decrease with increasing density, whereas torque density grows~\cite{maire2026kinetic}. Thus, in the relevant high-density, incompressible limit, torque-driven edge currents~\cite{Marconi2026hydrodynamics,metzger2026equationstateedgeflow} likely dominate the chiral response, with odd viscosity playing only a secondary role.

\begin{figure}
    \centering
    \includegraphics[width=0.9\linewidth]{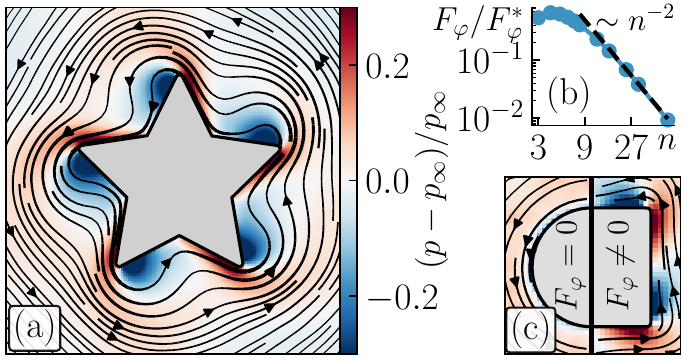}
    \caption{Chiral bath around a fixed intruder. (a) Pressure field $p$ approximated by half the trace of the virial stress and velocity streamlines for an achiral wheel in a chiral bath, with $R/r=2$ and $R/\sigma=20$. (b) Torque on regular $n$-gons with circumradius $R/\sigma=10$. (c) Same as (a) for a circle (left, $n\to\infty$) and a square (right, $n=4$). $\alpha_b=0.8$, $\Delta_b>0$, $n_b\pi\sigma^2=0.3$, $I\to\infty$, $M\to\infty$, $\Delta=0$ and $\alpha=1$. We define $\bar F_\varphi=\pi\tau R^2$. Each simulation ran for $10^8\sigma/\sqrt{T_b/m}$.}
    \label{fig:EDMD2}
    \vspace{-0.3cm}
\end{figure}
Fig.~\ref{fig:EDMD2}(a) shows the edge currents and pressure field around a fixed intruder in a chiral bath with $\Delta_b>0$. Circulation around the intruder produces pressure variations near the corners that support a steady flow without local accumulation. As shown in the companion paper~\cite{long}, this follows from a chiral Stokes equation differing from the achiral one only through the torque density $\tau$. Fig.~\ref{fig:EDMD2}(b) shows the torque $F_\varphi$ exerted by the bath on a regular $n$-gonal intruder ($n=3$ for a triangle, $n=4$ for a square, etc.). For $\Delta=0$, the collision rule enforces perfect slip, so no tangential traction is transmitted and the bath torque density $\tau$ does not act directly on the intruder. As a result, only the normal stress contributes. Neglecting viscous normal traction, we obtain:
\begin{equation}
    F_\varphi=\oint ds\big[\bm r_0\times\hat{\bm n}\big]_z\hat{\bm n}\cdot\mathsf\Sigma\cdot\hat{\bm n}\simeq-\oint ds\hspace{0.05cm}\big[\bm r_0\times\hat{\bm n}\big]_zp.
    \label{eq:torque_intruder}
\end{equation}
The intruder torque therefore arises not directly from the bath torque density, but from stress gradients generated by chiral edge currents, as illustrated in Fig.~\ref{fig:EDMD2}(a). Both mechanisms are absent in the dilute regime, which is instead controlled by explicit chirality in the bath-intruder interaction or by weak fluctuation-driven ratchet effects. Accordingly, Fig.~\ref{fig:EDMD2}(c) shows that for a circle the pressure around the intruder is uniform despite the circulating flow, so the torque vanishes $F_\varphi\to0$ as $n\to\infty$, whereas for a square the corners produce a nonuniform $p(s)$ and hence a finite torque. Using the chiral Stokes equation, we qualitatively capture both the nonmonotonic dependence of $F_\varphi$ on $n$ and its decay at large $n$ in the companion paper~\cite{long}. We can likewise show that the torque density and the chiral currents it generates induce an antisymmetric contribution to $\Upgamma$, \emph{even in the incompressible regime} if inertia is kept, and relate it to the breakdown of the Lorentz reciprocal theorem~\cite{long}. 

More broadly, the impact of torque density on odd response appears to have been largely overlooked~\cite{khain2024trading} and merits systematic investigation. A complete description of an intruder in a dense bath will also require a fluctuating hydrodynamic description, with possibly a kinetic boundary layer to capture ratchet effects arising from fluctuations and subtler odd responses such as those shown in Fig.~\ref{fig:EDMD}(a) and (c). 

\emph{Conclusion}---We have found a density-controlled crossover in the nonequilibrium chiral dynamics of an intruder. In the dilute regime, its behavior is captured by a Boltzmann-Lorentz description. As density increases, bath-intruder correlations and collective effects invalidate this theory, with transport coefficients that vanish in the dilute limit taking on non-zero values. This signals the breakdown of the kinetic description. At high density, the bath is better described as a continuum where chiral edge currents dominate the intruder's response. These results identify the distinct microscopic and hydrodynamic mechanisms governing chiral transport across density regimes.
\\
\textbf{Acknowledgments}: RM acknowledges useful discussions with Lorenzo Caprini, Umberto Marini Bettolo Marconi and Alessandro Petrini.\\
\onecolumngrid
\twocolumngrid

\bibliography{bib}

\clearpage
\onecolumngrid
\setcounter{page}{1}

\begin{center}
\Large{\textbf{Supplemental Material}}
\end{center}

\begin{comment}
% Reset and define supplement numbering
\setcounter{section}{-1}
\setcounter{subsection}{0}
\setcounter{subsubsection}{0}
\setcounter{secnumdepth}{3}

\renewcommand{\thesection}{S\arabic{section}}
\renewcommand{\thesubsection}{\thesection.\arabic{subsection}}
\renewcommand{\thesubsubsection}{\thesubsection.\arabic{subsubsection}}
\makeatletter

% Save the original sectioning commands
\let\suppoldsection\section
\let\suppoldsubsection\subsection
\let\suppoldsubsubsection\subsubsection

% Supplement-only table of contents
\newcommand{\supplementtoc}{%
  \begingroup
  \setcounter{tocdepth}{3}%
  \def\l@section{\@dottedtocline{1}{0em}{2.8em}}%
  \def\l@subsection{\@dottedtocline{2}{2.8em}{3.6em}}%
  \def\l@subsubsection{\@dottedtocline{3}{6.4em}{4.6em}}%
  \@starttoc{stc}%
  \endgroup
}

\renewcommand{\section}{%
  \@ifstar{\suppoldsection*}{\suppsection}%
}
\newcommand{\suppsection}[1]{%
  \suppoldsection{#1}%
  \addcontentsline{stc}{section}{\protect\numberline{\thesection}#1}%
}

\renewcommand{\subsection}{%
  \@ifstar{\suppoldsubsection*}{\suppsubsection}%
}
\newcommand{\suppsubsection}[1]{%
  \suppoldsubsection{#1}%
  \addcontentsline{stc}{subsection}{\protect\numberline{\thesubsection}#1}%
}

\renewcommand{\subsubsection}{%
  \@ifstar{\suppoldsubsubsection*}{\suppsubsubsection}%
}
\newcommand{\suppsubsubsection}[1]{%
  \suppoldsubsubsection{#1}%
  \addcontentsline{stc}{subsubsection}{\protect\numberline{\thesubsubsection}#1}%
}

\makeatother

\bigskip
\noindent{\large\textbf{Contents}}
\smallskip

\supplementtoc

\bigskip
\end{comment}

\section{Simulation}
\subsection{Monte-Carlo integration of the Boltzmann-Lorentz equation}

We simulate the rigid intruder in the dilute Boltzmann-Lorentz limit with an event-driven Monte Carlo algorithm. The intruder state is $\big(\bm X,\varphi,\bm V,\Omega\big)$, where $\bm X$ is the center-of-mass position, $\varphi$ the orientation, $\bm V$ the translational velocity, and $\Omega$ the angular velocity.

The intruder boundary is described in the body frame by a closed parametrization $\bm\rho_0(s)$ with $s\in[0,1]$. We discretize it into $N_s$ boundary elements,
\begin{equation}
s_i=\frac{i+1/2}{N_s}, \qquad i=0,\dots,N_s-1,
\end{equation}
and define the corresponding boundary points $\bm r_i=\bm\rho_0(s_i)$. For each element we precompute the unit tangent $\bm t_i$, the outward unit normal $\bm n_i$, the quadrature weight $w_i$, and the geometric factors
\begin{equation}
\tilde{\bm t}_i=\frac{\partial_s\bm\rho_0(s_i)}{|\partial_s\bm\rho_0(s_i)|}, \qquad w_i=|\partial_s\bm\rho_0(s_i)|\Delta s,
\end{equation}
with $\Delta s=1/N_s$, together with
\begin{equation}
\kappa_{n,i}=\big(\bm r_i\times\tilde{\bm n}_i\big)_z, \qquad \kappa_{t,i}=\big(\bm r_i\times\tilde{\bm t}_i\big)_z.
\end{equation}
The weights $w_i$ approximate the boundary line element, so that
\begin{equation}
\oint ds F(\bm r)  \simeq \sum_{i=0}^{N_s-1} w_i F(\bm r_i).
\end{equation}
At orientation $\varphi$, the body-frame vectors are rotated into the laboratory frame as
\begin{equation}
\hat{\bm n}_i(\varphi)=\mathsf R(\varphi)\cdot\tilde{\bm n}_i, \qquad \hat{\bm t}_i(\varphi)=\mathsf R(\varphi)\cdot\tilde{\bm t}_i.
\end{equation}

Between two collisions, the motion is ballistic,
\begin{equation}
\dot{\bm X}=\bm V, \qquad \dot\varphi=\Omega, \qquad \dot{\bm V}=0, \qquad \dot\Omega=0,
\end{equation}
so that over a free-flight time $\tau$,
\begin{equation}
\bm X\to\bm X+\bm V\tau, \qquad \varphi\to\varphi+\Omega\tau.
\end{equation}

For boundary element $i$, the contact-point velocity is
\begin{equation}
\bm U_i(\varphi,\bm V,\Omega)=\bm V+\Omega \hat{\bm z}\times\big(\mathsf R(\varphi)\cdot\bm r_i\big),
\end{equation}
and its normal component is
\begin{equation}
U_{n,i}(\varphi,\bm V,\Omega)=\bm U_i(\varphi,\bm V,\Omega)\cdot\hat{\bm n}_i(\varphi)=\bm V\cdot\hat{\bm n}_i(\varphi)+\Omega\kappa_{n,i}.
\end{equation}
Collisions are independent, and the collision rate attached to element $i$ is the incoming bath-particle flux through that element. The bath velocity distribution is \emph{assumed} Gaussian (only for numerical convenience, any distribution can be chosen instead),
\begin{equation}
f_{\mathrm M}(\bm v)=\frac{m}{2\pi T_b}\exp\biggl(-\frac{m|\bm v|^2}{2T_b}\biggr),
\end{equation}
and the local rate is
\begin{equation}
r_i(\varphi,\bm V,\Omega)=n_b w_i\int d\bm v \Theta\big[(\bm U_i(\varphi,\bm V,\Omega)-\bm v)\cdot\hat{\bm n}_i(\varphi)\big] (\bm U_i-\bm v)\cdot\hat{\bm n}_i f_{\mathrm M}(\bm v).
\end{equation}

Because the Maxwellian is isotropic, only the normal component $v_n=\bm v\cdot\hat{\bm n}_i(\varphi)$ enters after integrating over the tangential velocity. The projected distribution is Gaussian
\begin{equation}
\phi(v)=\frac{1}{\sqrt{2\pi}{v_{\rm th}}}\exp\biggl(-\frac{v^2}{2v_{\rm th}^2}\biggr),\qquad v_{\rm th}^2={T_b}/{m}.
\end{equation}
Therefore,
\begin{equation}
r_i(\varphi,\bm V,\Omega)=n_b w_i\int_{-\infty}^{U_{n,i}(\varphi,\bm V,\Omega)}\big(U_{n,i}(\varphi,\bm V,\Omega)-v\big)\phi(v) dv.
\end{equation}
Introducing the flux function
\begin{equation}
\Psi(U,{v_{\rm th}})=\int_{-\infty}^{U}(U-v)\phi(v) dv,
\end{equation}
we can write
\begin{equation}
r_i(\varphi,\bm V,\Omega)=n_b w_i\Psi\big(U_{n,i}(\varphi,\bm V,\Omega),{v_{\rm th}}\big).
\end{equation}
If we denote by
\begin{equation}
\chi(z)=\frac{1}{\sqrt{2\pi}}e^{-z^2/2}, \qquad H(z)=\int_{-\infty}^zdy \chi(y)
\end{equation}
the standard normal density and cumulative distribution, then
\begin{equation}
\Psi(U,{v_{\rm th}})={v_{\rm th}}\chi\biggl(\frac{U}{{v_{\rm th}}}\biggr)+U H\biggl(\frac{U}{{v_{\rm th}}}\biggr).
\end{equation}
The total collision rate is the sum over the boundary,
\begin{equation}
\nu(\varphi,\bm V,\Omega)=\sum_{i=0}^{N_s-1} r_i(\varphi,\bm V,\Omega).
\end{equation}
Because the collision rate varies continuously during free flight, the next collision time is sampled by Ogata thinning~\cite{ogata1981lewis}. We first construct an upper bound $\bar\nu$ of the instantaneous collision rate $\nu(t)$, draw candidate event times from a Poisson process of rate $\bar\nu$, and accept each candidate with probability $\nu(t)/\bar\nu$. The accepted events are then distributed according to the desired non-homogeneous Poisson process. More precisely, we have
\begin{equation}
\bar\nu(\bm V,\Omega)=\sum_{i=0}^{N_s-1} n_b w_i\Psi\big(|\bm V|+|\Omega||\kappa_{n,i}|,{v_{\rm th}}\big).
\end{equation}
A candidate waiting time is drawn from
\begin{equation}
\tau_{\mathrm{cand}}\sim \mathrm{Exp}(\bar\nu),
\end{equation}
and accepted with probability
\begin{equation}
\frac{\nu(\varphi+\Omega\tau_{\mathrm{cand}},\bm V,\Omega)}{\bar\nu}.
\end{equation}
Rejected candidates are discarded until a collision time $\tau$ is accepted.

After advancing the state to the accepted collision time, the colliding boundary element is selected with probability proportional to its local rate,
\begin{equation}
\mathbb P(i|\varphi,\bm V,\Omega)=\frac{r_i(\varphi,\bm V,\Omega)}{\nu(\varphi,\bm V,\Omega)}.
\end{equation}
Conditioned on a collision at element $i$, the incoming bath particle is sampled from the collision flux rather than from the bare Maxwellian. Writing $u=-g_n>0$ for the incoming normal relative speed, its conditional density is
\begin{equation}
p_i(u|U_{n,i})=\frac{u \phi(U_{n,i}-u)}{\Psi(U_{n,i},{v_{\rm th}})}=\frac{u}{\Psi(U_{n,i},{v_{\rm th}})}\frac{1}{\sqrt{2\pi}{v_{\rm th}}}\exp\biggl[-\frac{(u-U_{n,i})^2}{2{v_{\rm th}}^2}\biggr], \qquad u>0.
\end{equation}
This is the normalized flux integrand appearing in the rate. The incoming normal relative speed $u$ is generated by numerical inversion of its cumulative distribution. 

When the collision element of the intruder $i$ is chosen, we apply the collision rule:
\begin{equation}
\quad
\bm V'=\bm V+\frac{\bm J}{M},\quad
\Omega'=\Omega+\frac{(\bm r_0\times \bm J)_z}{I},\quad\bm v'=\bm v-\frac{\bm J}{m},
\label{eq:collision_rule}
\end{equation}
with
\begin{equation}
    J_t=\frac{2\Delta}{\lambda_t},\qquad J_n=\frac{1+\alpha}{\lambda_n}g_n-\frac{\kappa_n\kappa_t}{I\lambda_n}J_t.
    \label{eq:J}
\end{equation}
and start the prediction of another collision again.

In summary, each event of the simulation consists of four steps: sampling the next collision time by thinning, advancing the intruder ballistically to that time, selecting the colliding boundary element with weight $r_i$, and updating $\bm V$ and $\Omega$ after sampling the incoming normal relative speed from the flux-weighted law $p_i(u|U_{n,i})$. Repeating this procedure generates a continuous-time stochastic trajectory consistent with the discretized Boltzmann-Lorentz collision kernel used in the code.

\subsection{Event-driven molecular dynamics simulation of an intruder in a bath of hard particles}

We simulate both the bath and the intruder using event-driven molecular dynamics of hard particles. Because of the particles’ hard-core nature, a fully time-stepped algorithm cannot be used. Instead, we employ an event-driven algorithm in which the collision times for all relevant particle pairs are computed, and the system is then advanced to the earliest collision time. The corresponding collision is processed, after which the next set of collision times is recomputed, and the procedure is repeated iteratively.

The system is a box of size $L\times L$ with periodic boundary conditions. The time between collisions is computed with the help of a cell list and an event list as described in Ref.~\onlinecite{smallenburg2022efficient}.

\subsubsection{Bath-bath collision}

We first focus on the simulation of only bath particles. The bath particles are assumed to be circular, of mass $m$ and diameter $\sigma$. Two particles $i$ and $j$, with velocity $\bm v_i$ and $\bm v_j$, and position $\bm r_i$ and $\bm r_j$ at time $t=0$ collide after a time:
\begin{equation}
    \delta \tau_{ij}=\frac{-b - \sqrt{b^2 - \bm v_{ij}^2(\bm r_{ij}^2-\sigma^2)}}{\bm v_{ij}^2},
    \label{eq: coll time}
\end{equation}
with $b = \bm r_{ij}\cdot \bm v_{ij}$, $\bm r_{ij}=\Big[\bm r_i-\bm r_j\Big]_{\rm min.~image}$ and $\bm v_{ij}=\bm v_i-\bm v_j$. The minimum-image convention is used to compute distances. $\delta \tau_{ij}$ might be a complex number, in which case the collision never happens between this pair of particles. When the two particles collide ($|\bm r_{ij}|=\sigma$), their velocities change according to either a chiral collision rule:
\begin{equation}
\begin{split}
    \bm v_i'&= \bm v_i - \dfrac{1+\alpha_b}{2}(\bm v_{ij}\cdot \hat{\bm\sigma}_{ij})\hat{\bm\sigma}_{ij} +  \Delta_b\hat{\bm\sigma}_{ij}^\perp, \\
    \bm v_j'&= \bm v_j + \dfrac{1+\alpha_b}{2}(\bm v_{ij}\cdot \hat{\bm\sigma}_{ij})\hat{\bm\sigma}_{ij} - \Delta_b\hat{\bm\sigma}_{ij}^\perp,
\end{split}
\label{eq:coll_rule_chiral_bath}
\end{equation}
with $\hat{\bm\sigma}_{ij}=\bm r_{ij}/|\bm r_{ij}|$ and $\hat{\bm\sigma}_{ij}^\perp=\hat{\bm z}\times \hat{\bm\sigma}_{ij}=(-\hat \sigma_{ij, y}, \hat \sigma_{ij, x})$ or an achiral one:
\begin{equation}
\begin{split}
    \bm v_i'&= \bm v_i - \dfrac{1+\alpha_b}{2}(\bm v_{ij}\cdot \hat{\bm\sigma}_{ij})\hat{\bm\sigma}_{ij} +  \Delta_b^\parallel\hat{\bm\sigma}_{ij}, \\
    \bm v_j'&= \bm v_j + \dfrac{1+\alpha_b}{2}(\bm v_{ij}\cdot \hat{\bm\sigma}_{ij})\hat{\bm\sigma}_{ij} - \Delta_b^\parallel\hat{\bm\sigma}_{ij}.
\end{split}
\label{eq:coll_rule_non_chiral_bath}
\end{equation}
Both are used in the main text. Note that in both cases, the temperature of the bath $T_b$ is set by the balance between energy dissipated by $\alpha_b$ and injected by $\Delta_b$ or $\Delta_b^\parallel$~\cite{maire2026kinetic, brito2013hydrodynamic}.

To make the transport coefficients finite, we include a Langevin bath during the free-flight of some simulations. In this case, every particle $i$ receives a velocity update:
\begin{equation}
    \bm v_i\to\bm v_i-\gamma\bm v_i dt+\sqrt{2\gamma T_b dt/m}\bm \eta_i
\end{equation}
every $\gamma dt=1/10$. 
\subsubsection{Bath-intruder collision}

Each bath particle must also be checked against the rigid polygonal boundary of the intruder. During free flight, both the bath particle and the intruder move ballistically, namely
\begin{equation}
\bm r_i(\tau)=\bm r_i+\bm v_i\tau,
\qquad
\bm X(\tau)=\bm X+\bm V\tau,
\qquad
\varphi(\tau)=\varphi+\Omega\tau.
\end{equation}
At time $\tau$, the particle position relative to the intruder center is first reduced with the minimal-image convention and then rotated into the intruder body frame:
\begin{equation}
\bm q(\tau)=\mathsf R\big(-\varphi(\tau)\big)\cdot \Big[\bm r_i(\tau)-\bm X(\tau)\Big]_{\rm min.\ image}.
\end{equation}
If $\mathcal B_0$ denotes the intruder polygon in the body frame, we compute the signed gap
\begin{equation}
 g(\tau)=\text{sdist}\big(\bm q(\tau),    \mathcal B_0 \big)-\sigma/2,
\end{equation}
where $\text{sdist}$ is the signed distance to the polygon, positive outside and negative inside. Numerically, the unsigned distance is obtained as the minimum distance to all polygon edges, while the sign is determined by an even-odd point-in-polygon test. Therefore, the collision time with the intruder is the smallest positive root of
\begin{equation}
 g(\tau_c)=0,
\end{equation}
with $g(\tau)>0$ before contact.

To find the root, we search (by stepping with a small step $\delta \tau^{\rm step}$) for the first interval $[\tau_a,\tau_b]$ such that
\begin{equation}
 g(\tau_a)>0,
 \qquad
 g(\tau_b)\le 0.
\end{equation}
Once such a bracket is found, the collision time is refined by bisection until convergence up to a given tolerance. At this point, $\tau_a\simeq\tau_b$, but we use $\tau_a$ as the collision time, since it ensures no overlap.

The collision between the intruder and a bath particle is processed via the collision rule Eqs.~\eqref{eq:J} and \eqref{eq:collision_rule}. For a particle of finite size, the bath particle may collide exactly at the junction between two vertices of the polygonal intruder. In this case, we determine the closest point $\bm q$ on the polygon to the particle center $\bm p$, and define the contact normal as
\begin{equation}
    \hat{\bm n}=\frac{\bm p-\bm q}{|\bm p-\bm q|}.
\end{equation}
Therefore, if the closest point is the common vertex shared by two adjacent edges, the event is treated as a genuine vertex collision, and the normal is the radial direction from that vertex to the particle center.

\subsection{Numerical estimation of the effective Kramers-Moyal coefficients}

In both simulations, the Kramers-Moyal coefficients are not extracted from a coarse-grained time series sampled at a fixed time step. Instead, they are accumulated collision by collision, using the exact intruder's free-flight dynamics between two successive collisions and the impulsive jump at collision.  

\subsubsection{Body-frame versus Ornstein-Uhlenbeck}
Before describing how we estimate the Kramers-Moyal coefficient, we discuss an important point related to the difference between the Ornstein-Uhlenbeck process in the laboratory frame and the body frame.

The body frame itself rotates during a free flight.  Between two collisions, the laboratory velocities $(V_x,V_y,\Omega)$ are constant, but the body-frame translational components are not.  Indeed, going to the body-frame by rotating back at each time the intruder, we obtain:
\begin{equation}
\tilde{\bm V}(t)=\mathsf R\left[-\big(\varphi+\Omega t\big)\right]\cdot \bm V,
\end{equation}
and therefore the velocity evolution
\begin{equation}
\begin{pmatrix}
\dot{\tilde V}_x\\ \dot{\tilde V}_y\\ \dot \Omega
\end{pmatrix}_{\rm free}=
\begin{pmatrix}
\Omega\tilde V_y\\
-\Omega\tilde V_x\\
0
\end{pmatrix}.
\end{equation}
Therefore, because of the rotation of the intruder, the body-frame dynamics is not a plain Ornstein-Uhlenbeck process.  The effective body-frame equation instead has the structure:
\begin{equation}
\mathsf M\cdot\dot{\tilde{\bm U}}=\tilde{\bm F}-\tilde{\Upgamma} \cdot\tilde{\bm U}+
\begin{pmatrix}
M\Omega\tilde V_y\\
-M\Omega\tilde V_x\\
0
\end{pmatrix}+\tilde{\bm \eta}(t),\qquad \langle \tilde\eta_i(t) \tilde\eta_j(t')\rangle=\tilde{D}_{ij}\delta(t-t').
\label{eq:additional_force}
\end{equation}
This should be kept in mind when estimating the Kramers-Moyal coefficients, as we now describe.

\subsubsection{Change of velocity}
The change of momentum in the body-frame at collision $m$ is given by:
\begin{equation}
\Delta \tilde{\bm \Pi}_m^{\rm coll}=
\begin{pmatrix}
J_{b,x}\\
J_{b,y}\\
(\bm r\times\bm J_b)_z
\end{pmatrix},\qquad\bm J_b=\mathsf R(-\varphi_{m})\cdot\bm J.
\end{equation}

This collision, which happened at $t_m$, is followed by a free flight until the next collision at $t_{m+1}$. Between these collisions, the velocities in the laboratory frame are constant and written as $\bm U_m$ while the angle continuously changes with time as $\varphi_m + \Omega_m \Delta t$, where $\varphi_m$ is the intruder's angle at $t_m$.
 
To estimate the Kramers-Moyal coefficient, we will need the integrated translational velocity in the body-frame:
\begin{equation}
\int_{t_m}^{t_{m+1}}dt\tilde{\bm V}(t)=\int_0^{\Delta t_m}dt\mathsf R\left[-\big(\varphi_m+\Omega_m t\big)\right]\cdot\bm V_m ,\qquad \Delta t_m=t_{m+1}-t_m
\end{equation}
Writing
\begin{equation}
I_c=\frac{\sin(\varphi_m+\Omega_m\Delta t_m)-\sin\varphi_m}{\Omega_m},\qquad I_s=\frac{\cos\varphi_m-\cos(\varphi_m+\Omega_m\Delta t_m)}{\Omega_m},
\end{equation}
this gives
\begin{equation}
\int_{t_m}^{t_{m+1}}dt \tilde V_x(t) = V_{x,m} I_c + V_{y,m} I_s,\qquad\int_{t_m}^{t_{m+1}}dt \tilde V_y(t) = -V_{x,m} I_s + V_{y,m} I_c.
\end{equation}
Since $\Omega$ is constant during free flight, either in the body-frame or the laboratory frame, the integrated angular velocity is simply:
\begin{equation}
\int_{t_m}^{t_{m+1}}dt\Omega(t)=\Omega_m\Delta t_m.
\end{equation}

\subsubsection{Estimation of the Kramers-Moyal coefficients}

Each change of momentum at collision $\Delta\bm \Pi$ should be coarse-grained into a net force $\bm F$, a damping $-\mathsf{\Gamma} \cdot \bm U$, and a noise $\bm \eta$ with variance $\mathsf D$. That is, we need to find the best parameters such that:
\begin{equation}
\Delta\tilde{\bm \Pi}_m\approx\tilde{\bm F}^{\rm rot}\Delta t_m-\tilde{\Upgamma}\cdot\int_{t_m}^{t_{m+1}}dt\tilde{\bm U}(t)+\tilde{\bm \eta},\qquad\forall m.
\label{eq:deltapstimator}
\end{equation}
where
\begin{equation}
    \tilde{\bm F}^{\rm rot}=\tilde{\bm F}+\begin{pmatrix}
    M\langle \Omega\tilde V_y\rangle\\
    -M\langle \Omega\tilde V_x\rangle\\
    0
    \end{pmatrix},
\end{equation}
to account for the rotation of the body frame.

To formally obtain the best parameters, we rewrite Eq.~\eqref{eq:deltapstimator} as:
\begin{equation}
\bm y_m=\mathsf B^T\bm x_m+\bm r_m,
\end{equation}
with
\begin{equation}
\bm x_m=
\begin{pmatrix}
\Delta t_m\\
\int_{t_m}^{t_{m+1}}dt\tilde V_x(t)\\
\int_{t_m}^{t_{m+1}}dt\tilde V_y(t)\\
\Omega_m\Delta t_m
\end{pmatrix},\qquad \bm y_m=\Delta\tilde{\bm \Pi}_m,\qquad\mathsf B=
\begin{pmatrix}
\tilde{\bm F}^{{\rm rot},T}\\
\tilde{\Upgamma}^T
\end{pmatrix}.
\end{equation}
The residual $\bm r_m$ is the noise and will be evaluated using the deterministic dynamics. For now, we fit $\mathsf B$ by a  weighted least-squares fit,
\begin{equation}
\mathsf B_{LS}=\arg\min_{\mathsf B}\sum_m \frac{1}{\Delta t_m}\left\|\bm y_m-\mathsf B^T\bm x_m\right\|^2,
\end{equation}
with weights $w_m=1/\Delta t_m$. The first row of $\mathsf B_{LS}$ is the estimated force $\tilde{\bm F}_{LS}$, and the last three rows yield the matrix $\tilde{\Upgamma}_{LS}$. Both are evaluated in the body frame, and we recall that we did not take into account the intruder's rotation.

Since $\mathsf B_{LS}$ is known, we can obtain the residual:
\begin{equation}
\bm r_m=\bm y_m-\mathsf B^T_{LS}\bm x_m,
\end{equation}
whose variance estimates the diffusion matrix:
\begin{equation}
\tilde{\mathsf D}_{LS}=\frac{1}{T_{\rm obs}}\sum_m {\bm r}_m\otimes{\bm r}_m,\qquad T_{\rm obs}=\sum_m \Delta t_m.
\end{equation}

\end{document}